\documentclass[sigplan]{acmart}
\usepackage{fancyhdr}

\usepackage{enumitem}
\usepackage{graphicx}

\usepackage{fontawesome5}
\newcommand{\DatasetColor}{pink}
\newcommand{\ModelColor}{pink}
\newcommand{\InterfaceColor}{pink}
\newcommand{\DistributionColor}{orange}
\newcommand{\DeepfakeColor}{yellow}
\newcommand{\SearchEngineColor}{yellow}
\newcommand{\AdColor}{yellow}
\newcommand{\AppStoreColor}{yellow}
\newcommand{\DeveloperColor}{teal}
\newcommand{\CriticalColor}{teal}
\newcommand{\PaymentColor}{green}

\newcommand{\DatasetIcon}{database}
\newcommand{\ModelIcon}{code}
\newcommand{\InterfaceIcon}{magic}
\newcommand{\DistIcon}{comment}
\newcommand{\DFCIcon}{users}
\newcommand{\SearchIcon}{search}
\newcommand{\AppStoreIcon}{shopping-cart}
\newcommand{\DevIcon}{laptop-code}
\newcommand{\CriticalIcon}{tools}
\newcommand{\AdIcon}{bullhorn}
\newcommand{\PaymentIcon}{hand-holding-usd}

\newcommand{\iconlabel}[3]{%
\colorbox{#1!20}{%
\ \raisebox{-0.1ex}{\scalebox{0.8}{\faIcon{#2}}}\ #3\ %
}%
}
\newcommand{\Datasets}{\iconlabel{\DatasetColor}{\DatasetIcon}{Training Datasets}}
\newcommand{\GenerativeAIModels}{\iconlabel{\ModelColor}{\ModelIcon}{Generative AI Models}}
\newcommand{\GenerativeAIInterfaces}{\iconlabel{\InterfaceColor}{\InterfaceIcon}{Generative AI Interfaces}}
\newcommand{\DistributionChannels}{\iconlabel{\DistributionColor}{\DistIcon}{Distribution Channels}}
\newcommand{\DeepfakeCreationCommunities}{\iconlabel{\DeepfakeColor}{\DFCIcon}{Deepfake Creation Communities}}
\newcommand{\SearchEngines}{\iconlabel{\SearchEngineColor}{\SearchIcon}{Search Engines}}
\newcommand{\AdPlatforms}{\iconlabel{\AdColor}{\AdIcon}{Advertisement Platforms}}
\newcommand{\AppStores}{\iconlabel{\AppStoreColor}{\AppStoreIcon}{App Stores}}
\newcommand{\DeveloperPlatforms}{\iconlabel{\DeveloperColor}{\DevIcon}{Developer Platforms}}
\newcommand{\CriticalServiceProviders}{\iconlabel{\CriticalColor}{\CriticalIcon}{Critical Service Providers}}
\newcommand{\PaymentProcessors}{\iconlabel{\PaymentColor}{\PaymentIcon}{Payment Processors}}

\newcommand{\iconlabeltwo}[2]{%
\colorbox{#1!20}{%
\ \raisebox{-0.1ex}{\scalebox{0.8}{\faIcon{#2}}}\ %
}%
}

\newcommand{\IconGenerativeAIModels}{\iconlabeltwo{\ModelColor}{\ModelIcon}}
\newcommand{\IconGenerativeAIInterfaces}{\iconlabeltwo{\InterfaceColor}{\InterfaceIcon}}
\newcommand{\IconDistributionChannels}{\iconlabeltwo{\DistributionColor}{\DistIcon}}
\newcommand{\IconDeepfakeCreationCommunities}{\iconlabeltwo{\DeepfakeColor}{\DFCIcon}}

\newcommand{\IconAppStores}{\iconlabeltwo{\AppStoreColor}{\AppStoreIcon}}

\newcommand{\IconPaymentProcessors}{\iconlabeltwo{\PaymentColor}{\PaymentIcon}}

\AtBeginDocument{%
}

\copyrightyear{2026}
\acmYear{2026}
\setcopyright{cc}
\setcctype{by-nc-nd}
\acmConference[EAAMO '26]{Equity and Access in Algorithms, Mechanisms, and Optimization}{November 05--07, 2026}{Munich, Germany}
\acmBooktitle{Equity and Access in Algorithms, Mechanisms, and Optimization (EAAMO '26), November 05--07, 2026, Munich, Germany}
\acmDOI{10.1145/3846167.3848604}
\acmISBN{979-8-4007-3015-3/2026/11}

\begin{document}

\title[Mapping the Ecosystem of Technologies Facilitating AIG-NCII]{How to Stop Playing Whack-a-Mole: Mapping the Ecosystem of Technologies Facilitating AI-Generated Non-Consensual Intimate Images}

\author{Michelle L. Ding}
\email{michelle_ding@brown.edu}
\orcid{0009-0000-9778-1306}
\affiliation{%
\institution{Brown University}
\city{Providence}
\state{RI}
\country{USA}
}

\author{Harini Suresh}
\email{harini_suresh@brown.edu}
\orcid{0000-0002-9769-4947}
\affiliation{%
\institution{Brown University}
\city{Providence}
\state{RI}
\country{USA}
}

\author{Suresh Venkatasubramanian}
\email{suresh@brown.edu}
\orcid{0000-0001-7679-7130}
\affiliation{%
\institution{Brown University}
\city{Providence}
\state{RI}
\country{USA}
}

\renewcommand{\shortauthors}{Ding et al.}

\begin{abstract}
The last decade has witnessed a rapid advancement of generative AI technology that significantly scaled the accessibility of AI-generated non-consensual intimate images (AIG-NCII), a form of image-based sexual abuse that disproportionately harms and silences women and girls. There is a patchwork of commendable efforts across industry, policy, academia, and civil society to address AIG-NCII. However, these efforts lack a shared, consistent mental model that clearly situates the technologies they target within the context of a large, interconnected, and ever-evolving technological ecosystem. As a result, interventions remain siloed and are difficult to evaluate and compare, leading to a reactive cycle of whack-a-mole. In this paper, we contribute the first comprehensive AIG-NCII technological ecosystem that maps and taxonomizes 11 categories of technologies facilitating the creation, distribution, proliferation and discovery, infrastructural support, and monetization of AIG-NCII. First, we build and visualize the ecosystem through a synthesis of over a hundred primary sources from researchers, journalists, advocates, policymakers, and technologists. Then, we conduct two detailed walkthroughs to demonstrate the usefulness of the ecosystem in 1) making sense of new AIG-NCII harms using a case study of Grok and 2) mapping a clearer tech policy landscape using U.S. federal law and 63 state laws. We conclude with a vision for future AIG-NCII research that refines the edges of the ecosystem, recommending researchers to study critical relationships between technologies and potential ripple effects from different interventions. Our goal is to produce an AIG-NCII technological ecosystem that provides a clear, shared terminology and framework for stakeholders to move into the future of AIG-NCII prevention with clarity and foresight.
\end{abstract}



\maketitle

 \textcolor{red}{\textit{Content Warning: this paper includes discussions of image-based sexual abuse and gender-based violence, specifically AI-generated non-consensual intimate images.}}
 
\section{Introduction}

In December 2017, users on Reddit used a face-swap model to generate dozens of nude photos of women celebrities without their consent \cite{samanthacole_we_2018}. In December 2025, users on X used Grok AI to generate thousands of sexualized images of women and girls \cite{ceciliadanastasio_musks_2026}. In the eight years between these two cases, generative AI has evolved from simple face-swap models developed by independent programmers to huge multi-modal models owned by large AI companies, yet the gendered patterns of abuse have remained the same. How do we understand new incidents of harm and effectively design interventions to prevent abuse within a rapidly changing technology, policy, and advocacy landscape? 

\textbf{AI-generated non-consensual intimate images (AIG-NCII)}\footnote{A note on terminology: In this paper, we use the term \textit{AI-generated non-consensual intimate images (AIG-NCII)} to refer to the full spectrum of 1) non-consensual \textit{creation} of AI-generated intimate images, 2) non-consensual \textit{distribution} of AI-generated intimate images,  and 3) the act of threatening actions 1 or 2 for financial gain, also known as \textit{sextortion}. Others have used the acronym NCII to describe just the non-consensual \textit{creation} of intimate images but in this paper, our use of AIG-NCII covers the full spectrum of actions 1-3. We will occasionally use the acronym NDII to refer to just the non-consensual \textit{distribution} of intimate images. When AIG-NCII depicts children, it is also a form of \textit{AI-generated child sexual abuse material (AIG-CSAM)}. AIG-NCII has also been referred to as \textit{synthetic non-consensual explicit AI-created imagery (SNEACI)}. The media and general public have also colloquially referred to AIG-NCII as ``deepfake pornography,'' but activists have pushed against the use of this phrase because ``pornography'' falsely implies consent and legitimacy. The term ``child pornography'' has similarly been phased out because children cannot consent.} \textbf{are a form of image-based sexual abuse} that includes the non-consensual creation, distribution, and threat of creation or distribution of intimate images \cite{mcglynn_imagebased_2017}. AIG-NCII disproportionately depicts women and girls, making it a form of technology-facilitated gender-based violence \cite{kristinebaekgaard_technologyfacilitated_2024} that results in significant psychological, physical, financial, and reputational harm to victim-survivors \cite{mcglynn_its_2021,myimagemychoice_deepfake_2024,akerley_lets_2021} as well as a gendered chilling effect where victim-survivors retreat from online spaces due to fear of harassment \cite{flynn_deepfakes_2022,maddocks_deepfake_2020}. Generative AI significantly increases the accessibility of image-based sexual abuse and perpetuates a culture of non-consent at scale. The rapid advancement of generative AI technologies from small face-swapping models (e.g., DeepFaceLab \cite{timmerman_studying_2023}, DeepNude \cite{henryajder_state_2019}) to large multi-modal models (e.g., Stable Diffusion\cite{hawkins_deepfakes_2025a}, Grok \cite{emanuelmaiberg_elon_2025}) have lowered the barriers to image-based sexual abuse to the point that a single image of a person can be ``nudified'' in a matter of minutes \cite{mehta_can_2023,sophiecockerham_deepfake_2022,hawkins_deepfakes_2025a}. 

\textbf{AIG-NCII prevention is currently a game of whack-a-mole. } 
There have been many commendable efforts across industry, policy, academia, and civil society to address AIG-NCII, including U.S. federal law\footnote{This paper focuses on United States laws and policies, but the framework and recommendations are also applicable to other countries (see section \ref{sec:4.2}).} mandating the take-down of reported images \cite{sen.cruzted[r-tx]_take_2023}, almost 50 state laws with varying degrees of protections \cite{publiccitizen_share_2025}, local lawsuits against ``AI nudifier apps'' \cite{cityattorneyofsanfrancisco_city_2025,alexiosmantzarlis_taking_2025}, advocacy campaigns against distribution sites \cite{myimagemychoice_blockmrdeepfakes_}, open letters against payment processors and search engines \cite{nationalassociationofattorneysgeneral_legal_2025,nationalassociationofattorneysgeneral_legal_2025a} and AI safety techniques safeguarding generative AI models \cite{nist_reducing_2024}. However, these interventions target different technologies and use vague or inconsistent terminology (e.g., several U.S. state laws criminalize the creation of AI generated intimate images without a clear or consistent definition of \textit{who} among the AI supply chain counts as a ``creator'' \cite{publiccitizen_share_2025,kayleewilliams_us_2024}). 
What results is a fragmented patchwork of interventions that are difficult to evaluate, compare, and combine into a coherent, integrated approach. In practice, this leads to a reactive game of whack-a-mole: one image, model, or app gets taken down but another quickly pops into place. 

\textbf{AIG-NCII is powered not just by a single model, app, or website, but by a vast, interconnected, and ever-evolving technological ecosystem.} A more robust and effective strategy to combat AIG-NCII requires a clear understanding of the ecosystem of technologies that facilitate the creation, distribution, monetization, and proliferation of AIG-NCII. While prior work has studied pieces of this ecosystem (see section \ref{sec:relatedwork}), there is still a lack of a unified framework to understand the complex technological ecosystem as a whole. Our paper fills this gap: we synthesize over a hundred primary sources from academia, civil society, industry, and policy to build an AIG-NCII technological ecosystem that connects the dots and provides a clear, shared terminology and framework for stakeholders to move into the future of AIG-NCII prevention with clarity and foresight.

\textbf{We contribute the first comprehensive map of the technological ecosystem of AIG-NCII}, including 11 categories of technologies that span the \textbf{A) creation} (datasets, generative AI models, and generative AI interfaces), \textbf{B) distribution }(distribution channels), \textbf{C) proliferation and discovery }(deepfake creation communities, search engines, advertisement platforms, app stores), \textbf{D) infrastructural support} (developer platforms, critical service providers), and \textbf{E) monetization} (payment processors) of AIG-NCII. In section \ref{sec:relatedwork}, we outline the related work and methodology used to build out the AIG-NCII ecosystem. In section \ref{sec3}, we introduce the ecosystem and provide descriptions, examples, and history of the 11 technologies facilitating AIG-NCII paired alongside a visual map (Figure \ref{fig:map}). In section \ref{sec4}, we use two extensive walkthroughs to demonstrate the usefulness of the ecosystem in 1) making sense of new AIG-NCII harms and 2) mapping a clearer tech policy landscape. First, we conduct a close reading of five Grok-related news headlines to demonstrate how the ecosystem can be
used as a sense-making tool to build a cohesive understanding of the complex web of technologies that facilitate new AIG-NCII harms. Second, we manually review U.S. federal law and 63 state laws to demonstrate how the ecosystem can be used to map a clearer tech policy landscape. In both case studies, we show how the ecosystem is also robust over time as new articles get published or laws get passed. In section \ref{sec5}, we conclude with a vision for future AIG-NCII research that refines the edges of the ecosystem, recommending researchers to study critical relationships between technologies and potential ripple effects from different interventions. 

\section{Building the Ecosystem of Technologies Facilitating AIG-NCII}\label{sec:relatedwork}
In this section, we summarize the wide landscape of academic literature, journalistic reporting, technical documentation, civil society and advocacy resources, and U.S. federal and state law that informs our understanding of different parts of the broader AIG-NCII ecosystem. We organize these sources under four main tenets that guide our analysis.

\textbf{We situate AIG-NCII as a form of image-based sexual abuse and technology-facilitated gender-based violence.} First, we draw upon a large body of literature studying the prevalence and harms of image-based sexual abuse \cite{mcglynn_imagebased_2017,umbach_nonconsensual_2024,qin_did_2024,mcglynn_its_2021,mcglynn_revenge_2017,qiwei_sociotechnical_2024}, more specifically AI-generated non-consensual intimate images (AIG-NCII) of adults \cite{flynn_deepfakes_2022,maddocks_deepfake_2020,akerley_lets_2021,americansunlightproject_deepfake_2024} and children \cite{thorn_deepfake_2025,elizabethlaird_deep_2024,ciardha_ai_2025}. We also review existing research studying perceptions and attitudes towards ``deepfakes'' from school teachers \cite{wei_were_2025,elizabethlaird_deep_2024}, parents \cite{elizabethlaird_deep_2024}, youth \cite{thorn&weprotectglobalalliance_evolving_2024,elizabethlaird_deep_2024}, and the general public \cite{brigham_violation_2024,umbach_nonconsensual_2024,li_norms_2023,gamage_are_2022a,fido_celebrity_2022}. 

\textbf{We view AIG-NCII as an intersection of multiple disciplines and stakeholders.} In addition to research from gender \& feminism studies \cite{mcglynn_revenge_2017}, criminology \cite{flynn_deepfakes_2022,bates_revenge_2017}, porn studies \cite{maddocks_deepfake_2020,winter_deepfakes_2020a}, and social \& legal studies \cite{akerley_lets_2021,mcglynn_its_2021}, we also draw from the critical voices of on the ground advocates and practitioners. We reference  technical reports, open letters, and press releases from activists, coalitions, networks, and civil society organizations (e.g., Center for Democracy and Technology \cite{elizabethlaird_deep_2024}, Thorn \cite{thornalltechishuman_safety_2024,thorn&weprotectglobalalliance_evolving_2024}, StopNCII.org \cite{revengepornhelpline_stopncii_2025}, Cyber Civil Rights Initiative \cite{maryannefranks_ccri_2025}, American Sunlight Project \cite{americansunlightproject_deepfake_2024}, National Center on Sexual Exploitation \cite{patricktruemanesq._github_2023}, My Image My Choice \cite{myimagemychoice_deepfake_2024}, National Network to End Domestic Violence \cite{nationalnetworktoenddomesticviolence_techfacilitated_2024}, Rape, Abuse \& Incest National Network \cite{rainn_groks_2025}). In alignment with AI auditing literature that shows harms discovery occurs via journalism  \cite{birhane_ai_2024}, we synthesize news articles, audits, and technical reports from investigative journalists and trust \& safety professionals (e.g., the Indicator \cite{alexiosmantzarlis_ai_2025}, 404 Media \cite{emanuelmaiberg_elon_2025}, WIRED \cite{reecerogers_googles_2025}, Deeptrace \cite{henryajder_state_2019}, Tech Policy Press \cite{kayleewilliams_us_2024}, Graphika \cite{santiagolakatos_revealing_2023}, Bellingcat \cite{bellingcatsfinancialinvestigationsteam_faking_2025}) that have reported extensively on AIG-NCII in the past decade.

\textbf{We view AIG-NCII as product of ever-evolving technologies, including generative AI.} Thus, we draw from a rich landscape of socio-technical research that looks at specific pieces of the technological ecosystem, including distribution channels \cite{qiwei_reporting_2024}, deepfake creation communities \cite{timmerman_studying_2023,winter_deepfakes_2020a,widder_limits_2022}, AI ``nudifier'' and ``undressing'' apps \cite{gibson_analyzing_2024a,alexiosmantzarlis_ai_2025,williams_there_2025}, text-to-image models and developer platforms \cite{hawkins_deepfakes_2025a,wagner_perpetuating_2025}, the video generation ecosystem \cite{kamachee_video_2025}, and specific ``deepfake pornography'' distribution sites \cite{han_characterizing_2024}. We draw from AI safety research \cite{nist_reducing_2024} that looks at safeguarding models from producing AIG-NCII (e.g., filtering training datasets \cite{thiel_identifying_2024,tabone_pornographic_2021}, and keyword filtering for prompt-based models \cite{nist_reducing_2024}. We also recognize the technical and ethical limitations to these AI safety approaches, such as the ability for models to generate CSAM even after training data filtering \cite{okawa_compositional_2025,cretu_evaluating_2025} and the non-consensual use of nude images in deepfake research \cite{cintaqia_stop_2025}. In addition to academic literature, we reference industry practices by reviewing platform terms of service and acceptable use policies (e.g., from Github \cite{github_github_2025}, Civitai \cite{civitai_policy_2025}, and Reddit \cite{redditadmin_update_2018}). When relevant, we also examine technical artifacts (e.g., Github repositories for models \cite{available_upon_request} and datasets \cite{available_upon_request}).\footnote{In order to reduce traffic to technical tools that can produce AIG-NCII, we remove direct links to sensitive sources (e.g. NSFW datasets and repositories). Sources are available upon request (see Ethical Statement in section \ref{sec:ethics}).}

\textbf{We view AIG-NCII as a key concern for policymakers.} In order to make the ecosystem relevant for regulators, we manually reviewed U.S. federal and state law (e.g., the TAKE IT DOWN Act \cite{sen.cruzted[r-tx]_take_2023} and 63 ``intimate deepfake'' state laws as tracked by Public Citizen \cite{publiccitizen_share_2025});  lawsuits against ``AI nudifier apps'' \cite{cityattorneyofsanfrancisco_city_2025,alexiosmantzarlis_taking_2025} and letters to search engines and payment processors by the National Association of Attorneys General  \cite{nationalassociationofattorneysgeneral_legal_2025,nationalassociationofattorneysgeneral_legal_2025a}. Our systemic approach to mitigating the problems of AIG-NCII is further aligned with the recommendations of Wang et al \cite{wang_distinguishing_2026} in the broader setting of regulating generative AI systems. 

\textbf{Methodology.} Guided by these tenets, we iteratively analyzed and synthesized the heterogeneous sources described above into higher-level categories representing key technologies grouped by their role in supporting AIG-NCII. As we uncovered gaps in the emerging ecosystem (e.g., the role of developer platforms in hosting models and datasets), we actively sought out additional sources to fill in these gaps (e.g., technical artifacts, terms of service, and documentation pages of Github and Civitai). To produce a robust and evidence-based map, our focus was on technologies for which there are multiple concrete sources we could triangulate to characterize their role in the AIG-NCII ecosystem, rather relying on single sources or speculating about the possible roles of different technologies. Throughout our iterative mapping process, we aimed to achieve a level of granularity that was specific enough to differentiate distinct technical roles (e.g., splitting technologies into those that support discovery vs. monetization) and support the design of targeted interventions. At the same time, we also aimed to maintain a level of abstraction that could support a rapidly-changing ecosystem and interventions that recognize and target broader categories of technologies rather than individual models or apps. Section \ref{sec3} goes into further detail about sources used for each of the 11 technologies in the ecosystem. 
\begin{table*}[h]
\caption{Taxonomy of Technologies and their Roles in the AIG-NCII Ecosystem }
\label{tab:taxonomy}
\begin{tabular}{p{5cm} p{10cm}}
\toprule
Ecosystem Role & Technologies \\
\midrule
A. Creation & 
\Datasets 
\GenerativeAIModels
\newline
\GenerativeAIInterfaces\\

B. Distribution & 
\DistributionChannels\\

C. Proliferation and Discovery & 

\DeepfakeCreationCommunities \AppStores\newline
\SearchEngines 
\AdPlatforms
\\
D. Infrastructural Support 
& 
\DeveloperPlatforms
\CriticalServiceProviders 
\\
E. Monetization  & 
\PaymentProcessors 
\\
\bottomrule
\end{tabular}
\end{table*}
\section{Introducing the Ecosystem of Technologies
Facilitating AIG-NCII}\label{sec3}
In this section, we introduce the AIG-NCII ecosystem. We provide a taxonomy of 11 categories of technologies that play one of five roles in facilitating AIG-NCII: \emph{creation}, \emph{distribution}, \emph{proliferation and discovery}, \emph{supporting infrastructure}, and \emph{monetization}. Table \ref{tab:taxonomy} provides an overview of the 5 roles and the technologies associated with each role. Figure~\ref{fig:map} visualizes the ecosystem. The goal of our AIG-NCII ecosystem is not to reflect merely a snapshot in time. By identifying categories of technology as well as roles, we hope to depict a relatively stable ecosystem that can be a target for interventions, even as the underlying technologies evolve. 

\begin{figure*}[ht]
\centering
\includegraphics[width=.9\linewidth]{ding_etal_aigncii_ecosystem_diagram.jpg}
\caption{The ecosystem of technologies that facilitate AIG-NCII via \textbf{Creation} (training datasets, generative AI models, and generative AI interfaces), \textbf{Distribution}
(distribution channels), \textbf{Proliferation and Discovery} (deepfake creation communities, search engines, advertisement platforms, app
stores), \textbf{Infrastructural Support }(developer platforms, critical service providers), and \textbf{Monetization} (payment processors).}
\label{fig:map}
\Description{The ecosystem of technologies that facilitate AIG-NCII via 1) Creation (training datasets, generative AI models, and generative AI interfaces), 2) Distribution
(distribution channels), 3) Proliferation and Discovery (deepfake creation communities, search engines, advertisement platforms, app
stores), 4) Infrastructural Support (developer platforms, critical service providers), and 5) Monetization (payment processors).}
\end{figure*}

\textbf{A. CREATION} -- Technologies in this category (Training Datasets, Generative AI Models, Generative AI Interfaces) form a pipeline along the AI supply chain that enables the non-consensual \textit{creation} of AI-generated intimate images. 

\Datasets refer to text, image, video, audio, and/or multi-modal datasets used to train generative AI models. In the early 2010s face-swap models were trained using open-source face datasets and other curated NSFW datasets \cite{available_upon_request} comprised primarily of women. An early study of the 2019 DeepNude undressing app found that it was unable to generate images of men because it was trained only on images of women \cite{henryajder_state_2019}. Researchers have found that ``general-purpose'' image datasets like LAION-5B \cite{thiel_identifying_2024}, 
ImageNet \cite{prabhu_large_2020}, and LAION-400M \cite{birhane_ai_2024} also contain pornography, non-consensual intimate images, and/or known CSAM, enabling models that train upon them to generate AIG-NCII.

\GenerativeAIModels refer to a spectrum of models that use deep learning techniques to take in an input and produce a modified output \cite{nationalinstituteofstandardsoftechnology_glossary_2024}. In the 2010s, smaller, independent, \textbf{open-source models} like  DeepFaceLab \cite{available_upon_request}  primarily used general adversarial networks \cite{goodfellow_generative_2014,henryajder_state_2019} to automate the face-swapping process \cite{ding_malicious_2025}. DeepFaceLab in particular had direct ties to Mr.DeepFakes, a prolific ``deepfake pornography'' community and hosting site \cite{timmerman_studying_2023,han_characterizing_2024}. Early models required a high level of technical expertise and were inconsistent in quality and realism \cite{kamachee_video_2025}. However, the images they produced, despite being less realistic, were still extremely harmful to victim-survivors depicted and publicly documented and criticized by researchers, activists, and journalists \cite{samanthacole_we_2018,jamesvincent_ai_2017,brittneymcnamara_ai_2017,mcglynn_imagebased_2017}. With the 2020s came larger and more powerful generative AI models that produced realistic outputs. \textbf{Open-weight models} like Stable Diffusion and Flux have been fine-tuned to produce tens of thousands of \textbf{fine-tuned variants} designed to produce AIG-NCII, predominantly of women \cite{hawkins_deepfakes_2025a}. This trend is now being replicated with open-weight video-generation models \cite{kamachee_video_2025}. Additionally, proprietary \textbf{closed models} like xAI's Aurora, Google's Gemini models, and OpenAI's GPT models can be used via chatbot interface or API access to generate AIG-NCII \cite{reecerogers_googles_2025,jasonkoebler_groks_2026}. 

\GenerativeAIInterfaces increase the accessibility of AIG-NCII by providing an easy-to-use interface (e.g., website or app) for users to access generative AI model capabilities. In the 2010s face-swap era, there were already ``undressing'' apps like DeepNude that had over 95k active users \cite{henryajder_state_2019,samanthacole_we_2018}. However, in the 2020s with the advancement of generative AI technology, \textbf{AI nudifier applications} became a rapidly growing multi-million dollar economy \cite{alexiosmantzarlis_ai_2025,santiagolakatos_revealing_2023}  dedicated to the creation and monetization of AIG-NCII, predominantly of young women \cite{williams_there_2025,gibson_analyzing_2024a}. AI nudifier applications significantly lower the barrier to entry to creating AIG-NCII because any non-technical user can upload a photo of another person (like a yearbook photo or social media post \cite{samanthacole_massive_2025}) and create an ``undressed'' version within minutes. The apps offer features like undressing and positioning the subject in various sexual positions \cite{gibson_analyzing_2024a}. In addition to single-purpose AI nudifier applications, \textbf{general purpose model interfaces} have also been used to produce AIG-NCII. WIRED found that ChatGPT and Gemini have been used to ``strip women in photos down to bikinis'' \cite{reecerogers_googles_2025}. Most notably, Grok has been used to generate thousands of images of women and girls directly into the comment section of X \cite{ceciliadanastasio_musks_2026}. As AI companies begin adding ``spicy'' and ``erotica'' \cite{agencefrance-presse_openai_2025} modes into their chatbots, advocates are also sounding alarm bells for the predictable proliferation of AIG-NCII \cite{rainn_groks_2025}.

\textbf{B. DISTRIBUTION} -- 
%The technology in this category (Distribution Channels) encompasses hundreds of public platforms and private channels through which AIG-NCII is non-consensually disseminated once it is created. 
\DistributionChannels include both the \textbf{private channels} and \textbf{public platforms} that AI-generated intimate images may be non-consensually distributed across. A 2024 survey by the Center for Democracy \& Technology (CDT) of K-12 school students and teachers found that AI-generated NCII was most commonly shared through public channels (e.g., posting on social media platforms or adult sites) and private channels (e.g., direct message, text message, email) \cite{elizabethlaird_deep_2024}. These trends also align with traditional modes of non-consensual distribution of intimate images \cite{qin_did_2024}. Under the public platform category, there are also ``deepfake pornography sites'' such as Mr.Deepfakes that are dedicated to the distribution of AIG-NCII \cite{timmerman_studying_2023}. In 2024, the American Sunlight Project discovered tens of thousands of AIG-NCII depicting 26 senators and members of congress, 25 of them women, across eleven ``deepfake pornography'' websites  \cite{americansunlightproject_deepfake_2024}. Activists and researchers have called out hundreds of similar websites \cite{available_upon_request,myimagemychoice_blockmrdeepfakes_}. 

\textbf{C. PROLIFERATION AND DISCOVERY} -- AIG-NCII is not created and distributed in a vacuum. Technologies in this category (Deepfake Creation Communities, Search Engines, Advertisement Platforms, and App Stores) contribute to the widespread proliferation and discovery of technologies used to non-consensually create and distribute AI-generated intimate images. 

\DeepfakeCreationCommunities are online communities that provide general and technical assistance to members trying to create deepfakes. They are a ``key driving force behind the increasing
accessibility of deepfakes and deepfake creation software'' and serve as an ``entry point'' for new users to learn from experienced users \cite{henry_imagebased_2019}. Historically, ``deepfake creation communities'' have been large (reaching 100,000 members) and highly mobile (existing across platforms like Reddit, MrDeepFakes, 4chan, 8chan, Voat, Telegram, and Discord) \cite{henry_imagebased_2019,timmerman_studying_2023,winter_deepfakes_2020a}. In 2017, journalists uncovered ``r/deepfakes'' on Reddit, one of the earliest deepfake creation communities with over 90,000 subscribers \cite{adirobertson_reddit_2018} that focused on face swapping celebrity faces with porn performers \cite{samanthacole_we_2018,jamesvincent_ai_2017,brittneymcnamara_ai_2017}. In February, 2018, under public pressure, Reddit banned ``r/deepfakes'' and updated their site-wide rules ``against involuntary pornography and sexual or suggestive content involving minors'' \cite{redditadmin_update_2018,adirobertson_reddit_2018}. After the ``r/deepfakes'' ban, many members migrated to MrDeepFakes.com (MDF), the self-proclaimed ``largest deepfake community,'' which serves as both a deepfake distribution and consumption site and a forum for deepfake content creators \cite{timmerman_studying_2023,han_characterizing_2024}. MDF was also linked directly by the popular face-swap model ``DeepFaceLab'' on Github as a place to obtain technical support to use the model \cite{timmerman_studying_2023}. Researchers studying MDF found subforums on technical assistance for both models and datasets \cite{timmerman_studying_2023}. Most recently, Telegram group chats have served as ``deepfake creation communities'' for users using Grok AI to generate AIG-NCII \cite{emanuelmaiberg_telegram_2026}. 

\SearchEngines enable the wide-spread discovery of AI nudifier apps, AIG-NCII, and distribution sites. Independent researcher Genevieve Oh conducted 6000 Google searches of 100 public figures (elected officials, broadcasters, TV hosts, singers, and chess players) and found that 99.69\% of the searches for ``NAME+DEEPFAKE'' resulted in a ``deepfake pornography'' website result on the first page \cite{genevieveoh_dataset_2023}. Another audit found that queries for ``deepnude,'' ``nudify,'' and ``undress app'' on Google, Yahoo, and Bing all yielded at least one result leading the user to AI nudifier applications within the first 20 results \cite{chiarapuglielli_ecosystem_2025}. In 2025, 47 state attorneys general wrote an open letter to Google, Yahoo, and Microsoft urging them to block AIG-NCII content and creation tools from search engines. According to the letter, queries like ``how to make deepfake pornography,'' ``undress apps,'' ``nudify apps,'' or ``deepfake porn'' do not produce any any warning labels \cite{nationalassociationofattorneysgeneral_legal_2025a}. Instead, search engines ``quickly present users with direct links to deepfake NCII, to listicles rating the top apps for creating deepfake NCII, and to apps that enable the creation of naked and/or sexual images and videos of any person in any photo'' \cite{nationalassociationofattorneysgeneral_legal_2025a}. The AGs call for search engines to ``implement new policies that
appreciate the threat of deepfake NCII by steering users away from harmful content and
providing appropriate warnings'' \cite{nationalassociationofattorneysgeneral_legal_2025a}.

\AdPlatforms refer to platforms that enable organizations and individuals to place ads (specifically for AI nudifier apps) that drive massive traffic to these tools. AI nudifier apps gain users by posting advertisements on large social media platforms. Investigative journalists at 404 Media and the Indicator have documented thousands of advertisements that AI nudifiers post on Instagram \cite{emanuelmaiberg_instagram_2024,alexiosmantzarlis_meta_2025}. A technical report from Graphika tracking 34 AI nudifier apps demonstrates how these apps ``operate as a fully-fledged online industry''  and rely on advertising on mainstream social media platforms and deploying customer referral schemes such as referral links on platforms like Reddit and X, allowing them to accumulate over 24 million unique visitors in September, 2023 \cite{santiagolakatos_revealing_2023}. Graphika found a 2000\% increase in referral links in one year during 2023 \cite{santiagolakatos_revealing_2023}.

\AppStores are digital marketplaces for producers to upload and advertise apps and for consumers discover and download apps. The National Center on Sexual Exploitation has previously called out Apple and Google for their stores' role promoting facilitating sexual exploitation of children by allowing the discovery of inappropriate and potentially dangerous apps \cite{nationalcenteronsexualexplotation_dirty_2024,nationalcenteronsexualexploitation_apple_2023,bengoggin_canadas_2024}. Similarly, app stores enable the large-scale discovery and downloading of AI nudifier apps. According to Bellingcat, DeepSwap, an app that was featured on the top of Mr.DeepFakes, was available to Google Play and Apple stores \cite{bellingcatsfinancialinvestigationsteam_faking_2025}. With the most recent case of Grok, advocates have also critiqued app stores for continuing to host Grok and X \cite{carolinehaskins_why_2026,ronwyden_lujan_2026}. A 2026 report by the Tech Transparency Project found 55 apps in the Google Play Store that can ``digitally remove the clothes from women and render them completely or partially naked or clad in a bikini or other minimal clothing'' and 47 such apps in the Apple store \cite{techtransparencyproject_nudify_2026}. TTP found that these apps were ``downloaded more than 705 million times worldwide and generated \$117 million in revenue'' (a portion of which goes to Google and Apple). Both companies removed dozens of nudifier apps from their stores after press coverage \cite{murti_tech_2026,techtransparencyproject_nudify_2026}.

\textbf{D. SUPPORTING INFRASTRUCTURE} -- The creation, distribution, proliferation, and discovery of AIG-NCII relies on a vast digital infrastructure. Technologies in this category (Developer Platforms and
Critical Service Providers) provide critical infrastructural support that allows technologies on the creation and distribution levels to function. 

\DeveloperPlatforms are platforms for developers to create, store, manage, and share code, including models and datasets. Developer platforms are central to the development of open-source models and datasets that are used in the creation of deepfakes. Activists and researchers documented Github's role in hosting early face-swap models like DeepFaceLab, DeepNude, and Unstable Diffusion that were used to produce AIG-NCII \cite{winter_deepfakes_2020a,patricktruemanesq._github_2023,henryajder_state_2019}. AI-specific developer platforms like Hugging Face and Civitai are used to host open-weight models, such as pre-trained base models Stable Diffusion and Flux, as well as almost 35,000 fine-tuned ``deepfake'' variants, a majority of them sexual and signaling intent to produce non-consensual intimate images \cite{hawkins_deepfakes_2025a}. An audit of Civitai also found that it was predominantly used to host ``not-safe-for-work'' (NSFW) models and datasets \cite{wagner_perpetuating_2025}. After Civitai passed a new policy banning models depicting the likeness of real people in April 2025 \cite{emanuelmaiberg_civitai_2025}, 404 Media found that users downloaded over 5000 models and reuploaded them onto Hugging Face \cite{maiberg_hugging_2025}. 

\CriticalServiceProviders, including cloud service providers, domain name services (DNS), and authentication services, are support software, enabling the existence of multiple pieces of the technological ecosystem, including nudifier apps and deepfake websites. An Indicator audit of 85 AI nudifier websites \cite{alexiosmantzarlis_ai_2025} found that ``Amazon and Cloudflare provide hosting or content delivery services for 62 of the 85 nudifiers'' and ``Google enabled simple sign-on for 53 out of 85.'' On May 6, 2025, Mr.DeepFakes, the world's most notorious ``deepfake pornography'' site with over 650,000 users shut down because an unknown ``critical service provider terminated service permanently'' and ``data loss has made it impossible to continue operation'' \cite{alanawise_major_2025}. This case study demonstrates a connection between critical service providers and AIG-NCII distribution channels. 

\textbf{E. MONETIZATION} -- \PaymentProcessors (including credit cards and cryptocurrencies \cite{gibson_analyzing_2024a}) enable the monetization of non-consensual creation and distribution of AI-generated intimate images. The Indicator estimates the current AI nudifier economy of undressing apps to be over \$36 million dollars \cite{alexiosmantzarlis_ai_2025}. According to researchers, Mr.DeepFakes was used as an``actively growing deepfake market (primarily for people seeking to commission NSFW deepfake media)'' \cite{timmerman_studying_2023,han_characterizing_2024}. In 2025, 47 state attorneys general wrote a letter to Visa, Mastercard, American Express, PayPal, Google Pay, and Apple Pay stating ``sellers of deepfake NCII tools and content
have made their services available in exchange for fees paid via payment platforms...even including the
logos of those companies on their webpages'' and call for these companies to ``deny sellers
the ability to use their services when they are on notice of these connections but should be actively working to identify and remove any such sellers from their network'' \cite{nationalassociationofattorneysgeneral_legal_2025}. 
\begin{table*}[t]
\caption{Five Grok-related news articles and the technologies they discuss (1/7/26 - 1/9/26).}
\label{tab:grok}
\begin{tabular}{p{9cm} p{2.4cm}p{1cm}}
\toprule
News Headline and Annotations & Source & Date\\
\midrule
``Musk's \textbf{Grok AI} (\IconGenerativeAIInterfaces) Generated Thousands of Undressed Images Per Hour on \textbf{X} (\IconDistributionChannels)'' \newline & Bloomberg \cite{ceciliadanastasio_musks_2026} & 1/7/26 \\
``Inside the \textbf{Telegram Channel} (\IconDeepfakeCreationCommunities) Jailbreaking \textbf{Grok} (\IconGenerativeAIInterfaces) Over and Over Again'' \newline & 404 Media \cite{emanuelmaiberg_telegram_2026} & 1/7/26 \\
``Why Are \textbf{Grok} (\IconGenerativeAIInterfaces) and \textbf{X} (\IconDistributionChannels) Still Available in \textbf{App Stores} (\IconAppStores)?''\newline & WIRED \cite{carolinehaskins_why_2026} & 1/8/26 \\
``No, \textbf{Grok} (\IconGenerativeAIInterfaces) hasn’t \textbf{paywalled} (\IconPaymentProcessors) its deepfake image feature.''\newline & The Verge \cite{roberthart_no_2026} \newline & 1/9/26 \\
``Musk's \textbf{AI bot Grok} (\IconGenerativeAIInterfaces) limits some image generation on \textbf{X} (\IconDistributionChannels) after backlash'' & Reuters \cite{sriram_musks_2026}& 1/9/26 \\
\bottomrule
\end{tabular}
\end{table*}
\section{Usefulness of the AIG-NCII Ecosystem} \label{sec4}
In this section, we demonstrate the usefulness of the AIG-NCII ecosystem by walking through in detail how the ecosystem can be used to 1) make sense of new AIG-NCII harms via a case study of Grok and 2) map out a clearer tech policy landscape using U.S. federal law and 63 state laws. 

\subsection{Making Sense of New AIG-NCII Harms}\label{sec:grok}
In late December 2025, xAI's Grok chatbot produced thousands of sexualized images depicting adults and children, responding to waves of user prompts to ``put a bikini on her,'' and ``take her clothes off'' \cite{jasonkoebler_groks_2026,craigsilverman_briefing_2026,nanamgbechikwerenwachukwu_tfgbvgroknciidataset_2026,nickrobins-early_elon_2026}. Immediately, a surge of news followed with headlines like \textit{``Musk’s Grok AI Generated Thousands of Undressed
Images Per Hour on X''} \cite{ceciliadanastasio_musks_2026}, \textit{``Why Are Grok and X Still Available in App Stores?''} \cite{carolinehaskins_why_2026}, and \textit{``Inside the Telegram Channel Jailbreaking Grok Over and Over Again'' }\cite{emanuelmaiberg_telegram_2026}. When new incidents of AIG-NCII are discovered and reported along waves of headlines discussing triggering topics relating to image-based sexual abuse, the general public may find it difficult and overwhelming to make sense of how different technologies in the ecosystem are involved in facilitating harm. We use the case study of Grok to demonstrate how the ecosystem can be used as a referential tool for the general public to understand how different technologies intersect to cause new incidents of harm. 

\textbf{Step 1. Select news articles.}
We selected five Grok-related articles that were released between January 7th 2026 and January 9th 2026 in close succession from major news sources (Table \ref{tab:grok}).

\textbf{Step 2. Code the articles.}
Referencing the AIG-NCII ecosystem, we coded each of the article headlines with the technologies they discuss (see Table \ref{tab:grok}). 
\textbf{Step 3. Put together the analysis.}
All five articles mention ``Grok'' in the headline, but by using the ecosystem as a sense-making tool, we are able to break an all-encompassing ``Grok'' down into different technologies and gain a more informative analysis of the incident as follows:

\textit{Example Analysis. The incident of harm describes the chatbot ``Grok,'' a \textbf{generative AI interface} (\IconGenerativeAIInterfaces) that enables xAI users to directly access Grok (a family of \textbf{generative AI models} (\IconGenerativeAIModels) with the same name) \cite{ceciliadanastasio_musks_2026}. The chatbot Grok posts images directly onto the comment sections of X, a public \textbf{distribution channel} (\IconDistributionChannels) \cite{ceciliadanastasio_musks_2026} -- facilitating both the non-consensual creation and distribution of intimate images. This entire process is exacerbated by a \textbf{deepfake creation community} (\IconDeepfakeCreationCommunities), a Telegram channel consisting of thousands of users dedicated to jailbreaking Grok to produce AIG-NCII \cite{emanuelmaiberg_telegram_2026}. \textbf{App stores} (\IconAppStores) have also enabled users to download and access Grok \cite{carolinehaskins_why_2026}. On Jan 9, X restricted access to the Grok chatbot to paying subscribers, bringing in the role of \textbf{payment processors} (\IconPaymentProcessors) \cite{roberthart_no_2026}. Despite this policy, journalists found that free users can still access the \textbf{generative AI model} capabilities via other \textbf{generative AI interfaces} \cite{roberthart_no_2026}, including the grok.com website, the X platform, and the Grok mobile app for iOS and Android. This is because one \textbf{generative AI model} can have multiple \textbf{generative AI interfaces}. Limiting access to one (the chatbot) does not stop users from accessing others (website or app).}

\textbf{Step 4. Maintain a narrative over time by adding new information.}
AIG-NCII harms rarely only remain for one news cycle. In addition to allowing users to understand immediate headlines, the ecosystem can also serve as a tool to fit new information into the narrative. On April 14, 2026, NBC News reported that \textit{``Apple threatened to remove \textbf{Grok} (\IconGenerativeAIInterfaces) from the \textbf{App Store} (\IconAppStores) over sexualized deepfakes''} \cite{ingram_apple_2026}. After coding the headline using our ecosystem, we can see that after advocates called out the \textbf{app stores} for enabling the proliferation and discovery of Grok \cite{carolinehaskins_why_2026}, the \textbf{app stores} in turn put pressure on Grok, the \textbf{generative AI interface} used to create AIG-NCII. 

\textbf{Summary.} 
In this section, we describe a case study of applying the AIG-NCII ecosystem to Grok. We conduct a close reading of 5 related news headlines to demonstrate how the ecosystem can be used as a sense-making tool to build a cohesive understanding of the complex web of technologies that facilitate new AIG-NCII harms. 
Individual pieces of reporting and research necessarily focus on specific actors or technologies; the ecosystem connects these to each other and situates them within a broader technological landscape. 
We also show that the ecosystem's categories are robust enough to integrate new developments as they emerge, strengthening a running mental model of how these harms unfold over time.
\begin{table*}[h]
\caption{Table of U.S. Federal and State Laws that Target AIG-NCII Ecosystem Technologies (2023-2025)}
\label{tab:laws}
\begin{tabular}{p{1.6cm} p{2cm}p{11cm}}
\toprule
Jurisdiction & Legislation & Description \\
\midrule
Federal & 
S.146 (2023) & 

Prohibits non-consensual distribution of intimate images, both authentic and computer-generated, and requires \textbf{``covered platforms''} (\IconDistributionChannels) to remove reported depictions within 48 hours. \cite{sen.cruzted[r-tx]_take_2023} \newline \\
California & SB981 (2024) & Requires \textbf{``social media platforms''} (\IconDistributionChannels) to provide a mechanism that is reasonably accessible to a reporting user. \cite{californialegislativeinformation_sb981_2024} \newline \\
New York & A8808 (2024) & Enables individuals to maintain an action or special proceeding for a court order to require \textbf{``websites''} (\IconDistributionChannels) to permanently remove non-consensually distributed intimate images. \cite{newyorkA08808} \newline \\

Tennessee & H0769 (2025)  & 
Makes it an offense for a person to knowingly ``possess, distribute, or
produce \textbf{technology, software, or digital tools} (\IconGenerativeAIModels \IconGenerativeAIInterfaces) designed for the purpose of
creating material that includes a minor engaged in sexual activity or simulated sexual activity.'' \cite{tennesseegeneralassembly_hb_2025} \newline \\
Arkansas & HB1529 (2025) & Enables the Attorney
General to ``institute a civil action on behalf of the state against a \textbf{provider or developer of image generation technology} (\IconGenerativeAIModels\IconGenerativeAIInterfaces) that was used to create deepfake visual material.'' \cite{stateofarkansas_act_2025} \newline \\

Texas & SB 441 (2025) & Makes it an offense for an individual to non-consensually create intimate deepfakes. Does not apply to the ``\textbf{provider or developer of a publicly accessible artificial intelligence application or software} (\IconGenerativeAIModels\IconGenerativeAIInterfaces) that was used in the creation of the deep fake media'' \textit{if} they `took affirmative steps to prevent the creation of deep fake media.'' \cite{texasstatelegislature_sb_2025}\\
\bottomrule
\end{tabular}
\end{table*}

\subsection{Mapping a Clearer AIG-NCII Tech Policy Landscape}\label{sec:4.2}
\textbf{AIG-NCII is a crisis of non-consent at scale that regulators are scrambling to control.} While NCII has, unfortunately, existed long before generative AI entered the scene (e.g. ``upskirting'' or ``revenge pornography'') \cite{jenniferklein_call_2024}, \textit{AIG}-NCII has dramatically increased the speed and scale of image-based sexual abuse. Grok was generating over 6700 images per hour \cite{ceciliadanastasio_musks_2026}, and according to a 2026 report, nearly 90 schools and 600 students around the world have been impacted by AIG-NCII \cite{burgess_deepfake_2026}. The scale and prevalence of AIG-NCII contributes to a gendered chilling effect, where vulnerable communities retreat from online spaces due to fear of harassment \cite{flynn_deepfakes_2022,myimagemychoice_deepfake_2024,kristinebaekgaard_technologyfacilitated_2024,maddocks_deepfake_2020}. Schools in the Australia and South Korea have also provided students with the option to remove their photos from the yearbook for fear of being subject to deepfake abuse \cite{burgess_deepfake_2026}. As long as the technological ecosystem proliferates, the risk of abuse remains, and the silencing effect persists. Regulation that only criminalizes individuals for the creation, dissemination, or solicitation of AIG-NCII without targeting the \textit{technological ecosystem} supporting it is simply not enough to prevent abuse at scale.

\begin{figure*}[ht]
\centering
\includegraphics[width=0.65\linewidth]{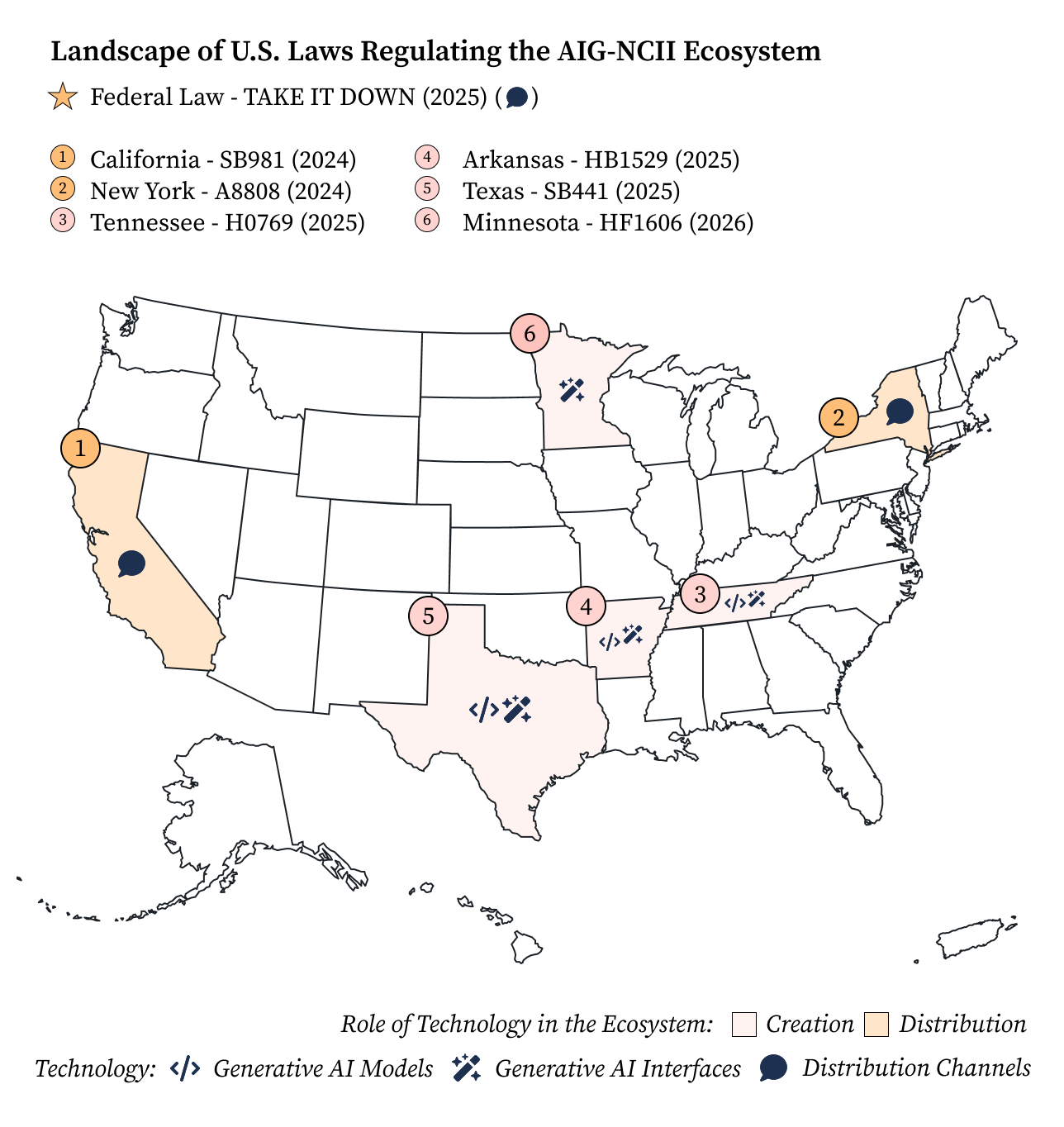}
\caption{The landscape of U.S. federal and state laws regulating the AIG-NCII ecosystem: \textbf{Federal Law} - The TAKE IT DOWN Act (2025); \textbf{California} - SB981 (2024); \textbf{New York} - A8808 (2024);  \textbf{Tennessee} - H0769 (2025); \textbf{Arkansas} - HB1529 (2025); \textbf{Texas} - SB441 (2025); \textbf{Minnesota} - HF1606 (2026). Each state is labeled with icons of regulated technologies and color-coded with the role of the technology in the ecosystem.}
\label{fig:law_map}
\Description{The landscape of U.S. federal and state laws regulating the AIG-NCII ecosystem. Laws included: Federal Law - The TAKE IT DOWN Act (2025);  California - SB981 (2024); New York - A8808 (2024); Tennessee - H0769 (2025); Arkansas - HB1529 (2025); Texas - SB441 (2025); Minnesota - HF1606 (2026). Each state is labeled with icons of regulated technologies and color-coded with the role of the technology in the ecosystem.}
\end{figure*}

\textbf{What does the current tech policy landscape look like?} In the U.S., there is a patchwork of federal and state laws that cover AIG-NCII to some degree. According to a legislative tracker by civil society organization Public Citizen, nearly all 50 U.S. states have some form of law addressing AIG-NCII  \cite{publiccitizen_share_2025}. However, these laws use inconsistent terminology, definitions, and include varying levels of protections for victim-survivors \cite{kayleewilliams_us_2024}. Here, we demonstrate how stakeholders in the tech policy space can use the ecosystem as a tool to build a clearer policy landscape by mapping laws in different jurisdictions to the technologies they target. In obtaining a clearer landscape, we can also conduct more intentional analyses that compare and contrast legal text across jurisdictions.

\textbf{Step 1. Select range of laws to map.} We manually read U.S. federal law (the TAKE IT DOWN Act \cite{sen.cruzted[r-tx]_take_2023}) and 63 state laws across 44 states\footnote{We last accessed the Public Citizen dataset on May 7, 2026 and filtered for bills that were marked as ``enacted'' (became state law) and that cover everyone.} \cite{publiccitizen_share_2025}. 

\textbf{Step 2. Use the ecosystem to code laws with technologies regulated.} Out of the 63 state laws, a vast majority target individuals who have created, solicited, or disseminated AIG-NCII. Here, the ecosystem helps us identify an overall gap in regulation: only 5 out of 63 state laws (7.9\%) explicitly target technology actors. For these 5 laws, we use the AIG-NCII ecosystem to manually code which technologies the text most likely regulates (Table \ref{tab:laws}).

\textbf{Step 3. Analyze the laws and their respective technologies.} Using Table \ref{tab:laws}, we 1) compare laws that target \textit{different} technologies, and 2) compare laws that target the \textit{same} technologies.

\textbf{3.1. Compare laws targeting \textit{different} technologies.} First, we can compare \textit{federal vs. state laws}. The only U.S. federal law to date targeting AIG-NCII is the TAKE IT DOWN Act \cite{sen.cruzted[r-tx]_take_2023}. Signed into law in May 2025 (and on effect in May 2026), the Act requires \textbf{public platforms} (\IconDistributionChannels) to remove reported non-consensually posted ``intimate visual depictions'' within 48 hours, to be enforced by the federal trade commission. When viewing the Act against our ecosystem, we can see that current U.S. federal protections only targets a very small piece of it, namely the \textbf{distribution channels}. When comparing federal law to the state laws listed in Table \ref{tab:laws}, we see that some states (California and New York) regulate the same piece of the ecosystem (\textbf{distribution channels}), while others (Tennessee, Arkansas, Texas) attempt to regulate technologies involved in the \textit{creation} of AIG-NCII (\textbf{generative AI models} and \textbf{generative AI interfaces}). 

We can further compare \textit{state vs. state} laws to understand the scope of protections across local jurisdictions. For example, we can compare Arkansas and California. Arkansas HB1529 (2025) \cite{stateofarkansas_act_2025} states that ``the Attorney General may institute a civil action...against a \textbf{provider or developer of image generation technology}  (\IconGenerativeAIModels\IconGenerativeAIInterfaces) that was used to create deepfake visual material'' if they did not have ``reasonable safeguards in place'' \cite{stateofarkansas_act_2025}, thereby targeting upstream \textbf{generative AI models} and \textbf{interfaces}. In contrast, California's SB-981 (2024) \cite{californialegislativeinformation_sb981_2024} would require a ``\textbf{social media platform} (\IconDistributionChannels)'' to provide California resident users with accessible means to ``report sexually explicit digital identity theft to the social media platform,'' i.e., targeting downstream \textbf{distribution channels} in a similar fashion to U.S. federal law. Here, we demonstrate how the ecosystem diagram and annotated laws (Table \ref{tab:laws}) provide a guide for intentional comparison of laws across jurisdictions that target \textit{different} technologies. 

\textbf{3.2. Compare laws targeting the \textit{same} technology.} Here, we compare the examples of Tennessee and Arkansas. TN H0769 (2025) targets actors who produce ``\textbf{technology, software, or digital tools} (\IconGenerativeAIModels\IconGenerativeAIInterfaces) designed for the purpose of creating material...'' \cite{tennesseegeneralassembly_hb_2025} whereas AK HB1529 (2025) specifically targets the ``\textbf{provider or developer of image generation technology} (\IconGenerativeAIModels\IconGenerativeAIInterfaces) that was to create deepfake visual material.'' \cite{stateofarkansas_act_2025}. While both laws reasonably target the creation pipeline, Tennessee's language focuses on how they are \textit{designed} while Arkansas' language focuses on how they have \textit{been used}. In practice, a \textbf{generative AI interface} that is clearly designed for the purpose of nudification (but never used) may not be subject to civil action on Arkansas but could be within scope in Tennessee. From this example, we see that although some jurisdictions target the same technology, they contain inconsistent language.

\textbf{Step 4. Maintain the tech policy landscape over time when new laws are enacted.} The AIG-NCII tech policy landscape is rapidly changing. Our ecosystem diagram is robust to these changes because it can also be used to continuously update the tech policy landscape over time when new laws are enacted. For example, on May 7, 2026, Minnesota passed a new state law (bill HF 1606) titled ``Prohibition on Nudification Technology'' \cite{__a}. Using the ecosystem, we can annotate the bill text with the technologies it targets (as done in Table \ref{tab:laws}): \textit{``A person who owns or controls a \textbf{website, application, software, program, or other service} (\IconGenerativeAIInterfaces) must not: 
(1) allow a user to access, download, or use the website, application, software, program, or other service to nudify an image or video; or
(2) nudify an image or video  on behalf of a user'' \cite{__a}}. Based on this annotation, Minnesota's new state law targets primarily \textbf{generative AI interfaces}, particularly nudifier applications. However, it differs from the other states that targets creation technologies (e.g. Tennessee and Arkansas) because \textbf{generative AI models} are not covered. 

\textbf{Summary.} In this section, we demonstrate, using a case study of U.S. federal law and 63 state laws, how the AIG-NCII ecosystem can be used as a tool to map out a tech policy landscape, assess the scope of individual laws, and compare and contrast laws across jurisdictions (see Figure \ref{fig:law_map}). We further show that our ecosystem is robust over time as new laws get passed. Mapping policies against the AIG-NCII ecosystem allows stakeholders to clearly characterize
and compare different state laws. As a result, industry players can improve product compliance; advocacy organizations can develop localized victim-survivor resources; lawyers have a larger toolkit for litigation; and state and local policymakers can situate their laws within a larger landscape. \textbf{This method can further be applied to international jurisdictions.} Governments around the world have recognized AIG-NCII of adults and children as a key synthetic content harm to urgently address \cite{nist_reducing_2024,governementsofaustraliacanadachiledenmarkspainfinlandfranceicelandmexicorepublicofkoreaunitedkingdomandsweden._global_2025,governmentofcanada_research_2025}. The AIG-NCII ecosystem can be used as a toolkit to understand how an international patchwork of laws can be used collectively to design more comprehensive, coherent interventions. 

\section{Refining the Edges: Future Directions for AIG-NCII Research}\label{sec5}
While prior work has investigated individual technologies contributing to AIG-NCII, the ecosystem surfaces the broader, interconnected technological landscape underlying these harms.
In doing so, it opens up new directions for future research that can in turn validate or inform refinements to the ecosystem.

\textbf{Studying the relationship between ecosystem technologies.} Our AIG-NCII technological ecosystem reveals edges (the relationships between technologies) that are important to understand but have not yet been studied. For example, to our knowledge, there has yet to be a study on the relationship between \textbf{generative AI models} and \textbf{generative AI interfaces} (specifically AI nudifier applications). Current AI nudifier apps or ``undressing'' sites state that they use generative AI technology, but do not specify the underlying model being used, reflecting a gap in tech accountability upstream: beyond the identified generative AI interface, we do not know exactly which generative AI models and training datasets are responsible for the resulting AIG-NCII. Further research on along this edge would assist existing interventions (e.g., lawsuits against the creators of these technologies) and shed light on future directions for AI governance (e.g., designing AI safety techniques to mitigate downstream harm). 

\textbf{Studying the ripple effects of interventions.} One implication of our ecosystem is that all the different technologies are connected in a complex web. There are many research opportunities around studying and leveraging ripple effects in the ecosystem to design better interventions. One such example is the possible ripple effects by \textbf{developer platforms}. For example, in April 2025, Civitai updated their Terms of Service to explicitly prohibit ``content that uses, reproduces, or is based on the likeness of real people - living or deceased - including public figures, celebrities, influencers, and private individuals, in any context'' \cite{civitai_terms_2025}. 404 Media has reported that Civitai's new policy ``deals [a] major blow to the nonconsensual AI porn ecosystem'' \cite{emanuelmaiberg_civitai_2025}. Using the AIG-NCII ecosystem map, researchers could study how this policy (compared to others put forth by Github, Hugging Face, or other developer platforms) impacts other technologies in the ecosystem, such as the amount of NSFW \textbf{generative AI models} or, further downstream, the number of \textbf{generative AI interfaces} available on \textbf{app stores}. Overall, we recommend researchers to adopt a relational approach to researching AIG-NCII by investigating the relationship between technologies in the ecosystem and the ripple effects that interventions may have. In doing so, we can collectively build up a stronger understanding of the ecosystem and how to design systems-level interventions. 

\section{Conclusion}
Image-based sexual abuse is a longstanding issue that predates generative AI. However, generative AI technologies and the vast technological ecosystem surrounding them enable abuse at scale, reinforcing the hypersexualization, dehumanization, and silencing of women and girls around the world. Taking down images, apps, websites, models, or datasets in an endless game of whack-a-mole is not enough to protect vulnerable populations within a rapidly evolving technological, advocacy, and policy landscape. In this paper, we contribute the first comprehensive ecosystem of 11 technologies that facilitate the  creation, distribution, proliferation and discovery, infrastructural support, and monetization of AIG-NCII through a synthesis of over a hundred sources from researchers, technologists, policymakers, journalists, trust and safety professionals, advocates, and civil society organizations (section \ref{sec:relatedwork}). We provide a taxonomy (section \ref{sec3}) and visualization (Figure \ref{fig:map}) of the ecosystem and conduct two extensive walkthroughs demonstrating how the ecosystem can be used to 1) make sense of new harms and 2) map out a clearer tech policy landscape (section \ref{sec4}). We conclude with a vision for future research in the AIG-NCII space, namely for researchers to study the edges of the ecosystem and potential ripple effects that interventions may have (section \ref{sec5}). Our goal is to provide a robust, clear, and accessible tool for stakeholders to collectively redirect of AIG-NCII prevention efforts away from recurring games of ``whack-a-mole'' and towards informed and collaborative efforts to eliminate this rapidly growing form of abuse.
\pagebreak
\bibliographystyle{ACM-Reference-Format}
\bibliography{references}
\newpage
\appendix
\section{Ethical Considerations}\label{sec:ethics}
In this section, we discuss ethical concerns or unintended adverse impacts that could result from the publication of our work, along with precautions we took taken to mitigate these issues.
\begin{enumerate}
\item \textbf{We did our best to prevent directing additional traffic to communities, tools, and platforms that facilitate AIG-NCII.} In building the ecosystem map, we had to cite technical tools, apps, websites, etc. that facilitated AIG-NCII. Malicious actors may use our paper as a starting point to discover these artifacts. In order to prevent further traffic to deepfake creation communities,  NSFW datasets/models, and other tools/software/websites, we attempted to first make our claims only by citing research papers, technical reports, and news articles discussing the apps rather than the link to the actual source (e.g. in section \ref{sec3}, we use a case study of DeepNude \cite{henryajder_state_2019} instead of actually citing the app). When alternative citation sources do not exist and we must reference artifacts (e.g. NSFW datasets or open source model repositories), we cite the placeholder ``Sensitive Primary Source. Available Upon Request.'' \cite{available_upon_request} (examples in section \ref{sec:relatedwork} and \ref{sec3}). Researchers can email us to receive a copy of these sources. Although this practice prevents immediate, open access to these sources, we believe that it is an important safety precaution to preserve. 
\item \textbf{We do not engage in methods that may result in engaging with illegal content (e.g. child sexual abuse material) or unethical actions (e.g. producing AIG-NCII).} Throughout the entire research and writing process, we did not engage in any capacity with actual AIG-NCII content depicting children or adults. The entire ecosystem is purely built upon research, news articles, reports, open letters, technical artifacts, and technical reports. We never downloaded or used any models, apps, or websites to produce AIG-NCII. We believe that this limitation did not hurt the quality of our ecosystem diagram. 
\item \textbf{We use survivor-centered terminology that aligns with existing advocacy around AIG-NCII.} We recognize that the language in research papers may reinforce harmful terminology and perspectives around sexual violence. We take care to use terminology that aligns with existing advocacy efforts around image-based sexual abuse. For example, our choice to use ``AI-generated non-consensual intimate images'' or ``image-based sexual abuse'' instead of the common phrase ``deepfake pornography'' aligns with existing calls by advocates against using the word ``pornography,'' which falsely suggests consent and legality \cite{mcglynn_not_2016,internetwatchfoundation_child_2025}. 

\item \textbf{We consider the possibility that exposing limitations in regulation could enable malicious actors to circumvent liability.} In section \ref{sec:4.2}, we demonstrate how the ecosystem can be used as a tool to map out the U.S. federal and state legal landscape around AIG-NCII. Such a clear landscape is useful for tech policy stakeholders to improve prevention efforts. In reverse, it may also be a useful guide for malicious actors to circumvent existing regulation. However, we believe that the benefits outweigh the consequences in this particular discussion. We are also actively seeking to share our findings with policymakers and others in the tech policy space to make them aware of these inconsistencies. 
\end{enumerate}
Overall, we took precautionary measures to limit the ethical concerns and adverse impacts that our paper could bring. We believe that our ecosystem makes important strides in bridging gaps across disciplines and stakeholders and advancing AIG-NCII prevention to protect vulnerable communities online. 

\end{document}